\documentclass[12pt,a4paper]{paper}
\usepackage{amsmath}
\usepackage{amsfonts}
\usepackage{amssymb}
\usepackage{textcomp}
\usepackage{graphics}
\usepackage[latin1]{inputenc}
\usepackage{natbib}
\usepackage{graphicx}

\title{A numerical study of the longitudinal thermoconvective rolls in a mixed convection flow in a horizontal channel with a free surface}
\author{Lahcen BAMMOU$^{\dag\ddag}$  \and Kamal EL OMARI$^{1\dag}$ \and Serge BLANCHER$^{\dag}$ \and Yves LE GUER$^{\dag}$  \and Brahim BENHAMOU$^{\ddag}$ \and Touria MEDIOUNI$^{\maltese}$}

\institution{${\dag}$Laboratoire des Sciences de l'Ingénieur Appliquées à la Mécanique\\ et au génie Électrique (SIAME), Fédération IPRA-CNRS, \\Université de Pau et des Pays de l'Adour (UPPA),\\ Bat. d'Alembert, Avenue Jules Ferry, 64075 Pau Cedex, France. \and 
${\ddag}$Laboratoire de Mécanique, Procédés de l'Energie\\ et de l'Environnement (LMP2E),\\ ENSA, B.P. 1136 -  80000 Agadir - Maroc \and 
${\maltese}$Laboratoire de Mécanique des Fluides\\ et d'Énergétique (CNRST-URAC27), \\
Université Cadi Ayyad - Dépt. de Physique,\\ Faculté des Sciences Semlalia, Marrakech 40001, Maroc.}

\begin{document}
\maketitle
\footnotetext[1]{Corresponding author: kamal.elomari@univ-pau.fr\\ Preprint submited to International Journal of Heat and Fluid Flow}

\begin{abstract}
This paper presents a numerical study of three-dimensional laminar mixed convection within a liquid flowing on a horizontal channel heated uniformly from below. The upper surface is free and assumed to be flat. The coupled Navier-Stokes and energy equations are solved numerically by the finite volume method taking into account the thermocapillary effects (Marangoni effect). When the strength of the buoyancy, thermocapillary effects and forced convective currents are comparable $(Ri\backsimeq O(1)$ and $Bd=Ra/Ma \backsimeq O(1))$, the results show that the development of instabilities in the form of steady longitudinal convective rolls is similar to those encountered in the Poiseuille-Rayleigh-Bénard flow. The number and spatial distribution of these rolls along the channel depend on the flow conditions. The objective of this work is to study the influence of parameters, such as the Reynolds, Rayleigh and Biot numbers, on the flow patterns and heat transfer characteristics. The effects of variations in the surface tension with temperature gradients (Marangoni effect) are also considered.
\end{abstract}


\section{Introduction}
The study of mixed convection in a horizontal channel with a free upper surface is of considerable interest for many industrial applications, such as heat exchangers, evaporative cooling devices, and chemical vapor process. In particular conditions, the resulting buoyancy forces induce the development of longitudinal vortex rolls, which constitute the first stage of transition to turbulence and contribute to an enhancement of heat transfer \citep{imura1978}.\\
Several analytical, numerical and experimental analyses involving mixed convection heat transfer in different channel geometries have been reported in the literature. However, it is common practice to model a thin liquid film on a wetted wall as a boundary condition for the heat and mass transfer in the adjacent gas flow. Thus, laminar mixed convection heat and mass transfer in wetted vertical ducts has been analyzed by \cite{lin} and \cite{tsayandyan}. They showed that the combined thermal and mass diffusion buoyancy forces have a great effect for small Reynolds number flows with high inlet temperatures. \cite{Oulaid2010} conducted a similar study that analyzed the effect of flow reversal on inducing flow instability. When the thickness of the liquid film is not negligible, it became interesting to study the flow mechanisms inside the film itself.

In this study, we consider the case of mixed convection inside a liquid film with a free surface where thermocapillary effects are present. When a free liquid surface is present, the surface tension variations resulting from the temperature gradient along the surface may induce motion within the fluid, called thermocapillary flow (i.e., thermal Marangoni convection). To the best of our knowledge, the combination of buoyancy, thermocapillary forces and forced axial flow has not been addressed in the literature. However, this original flow situation is the combination of other flow situations (of physical effects) that has been extensively studied. One of these situations is the flow inside a horizontal channel with the superposition of a forced axial flow and vertical natural convection currents. This unstable flow situation is known as the Poiseuille-Rayleigh-Bénard (PRB) flow and has flow structures that are similar to the flow along an open channel that we are interested in studying.

Many studies, both theoretical and experimental, have been conducted on mixed convection heat transfer in horizontal channels with rigid upper surface. The onset of a secondary flow, in the form of longitudinal rolls, takes place around a critical Rayleigh number of $Ra_c=1708$ \citep{akiyama1971,yasuo1966} for which the roll diameter is nearly equal to the channel height. \cite{ostrach1975} performed experiments for fully developed air flow between isothermal plates and noted that the vortex rolls become irregular for $Re=38$ and $Ra>8000$. \cite{Nicolas2000} presented the linear stability analysis of a fully developed Poiseuille flow in a duct and established a stability diagram. \cite{Bonnefoi2004} investigated the thermoconvective instabilities for a mixed convection water flow in a horizontal rectangular duct uniformly heated from below. The authors distinguish several regimes of flow and, specifically, the domains of thermoconvective instability. \cite{Wang2003} investigated three-dimensional mixed convection flow and heat transfer in a horizontal duct using constant or variable thermophysical property models. The numerical results from the two different approaches have been compared. \cite{Luijkx1981} demonstrated the existence of unsteady transverse thermoconvective rolls with the axis perpendicular to the main flow for the flow of silicone oil ($Pr=450$). For low Reynolds numbers, the critical Rayleigh number corresponding to the onset of the transverse rolls was found to be a function of the aspect ratio and the Prandtl number. A transition from transverse to longitudinal roll flow structure was found, and the corresponding boundary in the (Re-Ra) plane was experimentally established by \cite{Ouazzani1990,Ouazzani1993} and numerically by \cite{nicolas1997}. A comprehensive literature review dedicated to the Poiseuille-Rayleigh-Bénard flow structure was presented by \cite{nicolas2002}.

A second flow situation that is similar to the one studied in this paper is the case of a channel whose upper surface is free but without thermocapillary forces. This case has been less studied. Experiments were performed by \cite{gilpin1978} to confirm the occurrence and growth of longitudinal vortices in the laminar boundary layer developing in the water over a heated horizontal flat plate with a uniform surface temperature. Temperature profiles across the boundary layer were measured for flows with and without vortices to show qualitatively the effect that longitudinal vortices have on the heat transfer rate on the plate. An experimental study was also performed by \cite{wang1983} to determine the hydrodynamic and thermal conditions of water flow in an open channel that is uniformly heated from below. The authors also used a two-dimensional boundary layer model to predict the flow behavior.

Flow situation studies, where thermocapillary forces (Marangoni convection) have been taken into account, are essentially related to flows without forced convection. These flows have been extensively studied in the literature \citep{Goncharova2010, Simanovskii2008, Sadiq2010, MAO2008, Medale2009}. A pioneer study of thermocapillary instabilities was conducted by \cite{Pearson1958} who analyzed the stability of a horizontal layer with a non-deformable free surface. The liquid phase is modeled using the classical fluid mechanics equations, while the gas phase is included in the model through a boundary condition at the liquid-gas interface, which is characterized by a Biot number. Through a linear stability analysis, Pearson identified the existence of a critical threshold and determined a critical Marangoni number from which the liquid layer becomes unstable due to the thermocapillary convective motion that appears spontaneously. \cite{Scriven1964} extended Pearson's study by considering a deformable liquid-gas interface. Unlike the situation addressed by Pearson, they concluded that the liquid layer is still unstable and there is no critical Marangoni number. More recently, the coupling between thermocapillary and buoyancy has been considered through different approaches. To verify the influence of buoyancy forces, \cite{Laure1989} conducted several studies for low Prandtl number ($Pr$) fluids, while \cite{Parmentier1993} studied liquids with $Pr$ up to 10. \cite{burguete2001} found that thermocapillary-buoyancy convection could destabilize the flow into different patterns, depending on the temperature difference $\Delta T$ and on the liquid pool depth $d$. For small $d$ values, the system exhibits hydrothermal waves, while for larger $d$ values, stationary longitudinal rolls are observed. The stability of buoyant-thermocapillary instabilities has been analyzed by \cite{Mercier1996}. They introduce a Biot number to describe the heat transfer at the top free surface and showed the existence of a transition from oscillatory to stationary modes when the Bond number (the ratio between the Rayleigh and the Marangoni numbers) is increased. This transition depends on the Biot number.

Other studies have examined a thin liquid film flowing over an inclined heated plate, taking into account only the thermocapillary effect \citep{sadiq2005,thiele2004}. In these studies, most authors did not consider the thermoconvective instabilities that occur within a liquid film uniformly heated from below. Indeed, when the strength of buoyancy, thermocapillary and forced convective currents are comparable, $(Ri\backsimeq O(1)$ and $Bd=Ra/Ma \backsimeq O(1))$, and the instabilities develop in the form of longitudinal convective rolls similar to those found in Poiseuille-Rayleigh-Bénard.

This study is dedicated to the numerical simulation of three-dimensional laminar mixed convection of a liquid in a horizontal channel uniformly heated from below with a free upper surface. The objective is to study the influence of the control parameters (i.e., Reynolds, Rayleigh and Biot numbers) on the heat transfer in the liquid. The effects of the surface tension and their variations with temperature gradients (Marangoni effect) are taken into account. The paper is organized as follows. The physical model and governing equations are first described, followed by a description of the numerical resolution method and code validation. The effect of the studied parameters on the longitudinal rolls and heat transfer are then assessed. Finally, the impact of the coupled effects of natural convection and Marangoni convection is analyzed. Conclusions are then drawn based on these results.

\section{Physical model}
\begin{figure}[h!]
  \begin{center}
\includegraphics[height=3.3cm , width=0.60\textwidth]{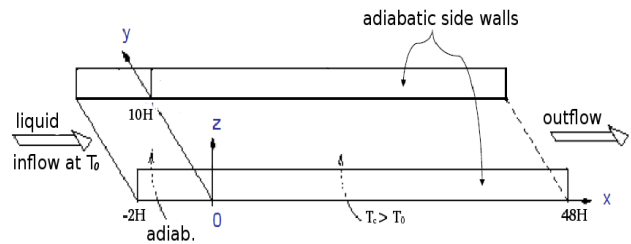} 
\caption{\it \small The studied configuration and corresponding boundary conditions.}\label{figure1}
  \end{center}
\end{figure}

We consider a liquid flow ($Pr = 7$) in a horizontal rectangular channel of height $H$, width $10H$ and length $50H$ with a free upper surface (Figure \ref{figure1}). We define the transverse aspect ratio as $\Gamma$ = width/height = 10. The side walls of the channel are adiabatic. To avoid the formation of the convective rolls immediately at the inlet section, we consider an adiabatic entrance over a length of $2H$. Beyond this adiabatic entrance, the bottom-wall is kept at a constant and uniform temperature $T_c$, which is higher than that of the inlet liquid $T_0$ ($T_c> T_0$). The temperature of the ambient air over the liquid flow is fixed at $T_0$.

The fluid is assumed to be Newtonian and incompressible and undergoes steady laminar flow. The thermophysical properties are assumed to be constant except for the density in the buoyancy term and the surface tension. The Boussinesq assumption is assumed, with $\rho$ taken to be constant everywhere, except in the buoyancy term, where a linear law is considered: $\rho=\rho_0(1-\beta(T-T_0))$, where $\beta$ is the thermal expansion coefficient and $\rho_0$ and $T_0$ are the reference values for the density and temperature, respectively. This assumption is justified for a sufficiently weak temperature difference across the fluid \citep{Gray1976545}. The variation of the surface tension with temperature is modeled by the following linear approximation: $\sigma=\sigma_0-\gamma(T-T_0)$, where $\sigma_0$ is the surface tension at temperature $T_0$ and $\gamma=-\partial
\sigma/\partial T$. The free surface is assumed to be flat and is subject to a convective heat exchange with ambient air characterized by a heat transfer coefficient $h$. The surface deformation can be neglected if the Crispation number $Cr=\rho\nu\alpha/\sigma H \ll 1$ \citep{Davis}, where $\rho$, $\nu$ and $\alpha$ denote the fluid density, kinematic viscosity and thermal diffusivity, respectively. Moreover, the Galileo number $Ga=gH^3/\nu\alpha \gg 1$ characterizes the relative importance of gravity and diffusion ($g$ is the gravity constant). A large value of the Galileo number indicates that gravity stabilizes the long-wave mode of the free surface deformation. Such conditions are shown be satisfied for the problem considered here, where
$Cr=1.42\times10^{-13}$ and $Ga=7\times10^{13}$ for $Ra=5000$, $Re=15$, $Bi=15$ and $Ma=500$. Using these assumptions, the equations governing the conservation of mass, momentum and energy are written as follows:

\begin{eqnarray}
 \vec{\nabla}.\vec{v}&=&0 \label{eq:continuite}\\
(\vec{v}.\vec{\nabla})\vec{v}&=&-\frac{1}{\rho_0}\vec{\nabla}
p+\nu\vec{\nabla^2}\vec{v}+\beta\vec{g}(T-T_0)\label{eq:conservation}\\
\vec{v}.\vec{\nabla}T&=&\alpha\vec{\nabla^2}T \label{eq:energie}
\end{eqnarray}
\\
Where $\vec{v}$ denote the three-dimensional velocity field and $T$ is the temperature field.

At the bottom of the channel, the velocity satisfies the no-slip condition, and the wall is assumed to be adiabatic for $x\in[-2H,0[$ and isothermal ($T=T_c$) for $x\in[0,48H]$:

\begin{equation}
  \label{eq:bottom-cond}
  \vec{v}=\vec{0},\hspace{0.4cm} \frac{\partial T}{\partial z}=0 ~~\text{for}~ x\in[-2H,0[;\hspace{0.4cm} T=T_c ~~\text{for}~
x\in[0,48H] \hspace{0.4cm} \text{at}~ z=0.
\end{equation}

The velocity satisfies also the no-slip condition for the side walls, which are assumed to be adiabatic:

\begin{equation}
  \label{eq:side-cond}
  \vec{v}=\vec{0},\hspace{0.5cm} \frac{\partial T}{\partial y}=0 \hspace{0.5cm} \text{at}~ y=0~~\text{and}~~y=10H.
\end{equation}

By assuming a planar surface, a shear stress boundary condition is imposed at the free surface, derived from the balance between the surface tension forces and the viscous stresses in the fluid:

\begin{equation}
  \label{eq:shear-stress}
\mu\frac{\partial v_x}{\partial z}=-\frac{\partial\sigma}{\partial T}\frac{\partial T}{\partial x},\hspace{1cm} \mu\frac{\partial
v_y}{\partial z}=-\frac{\partial\sigma}{\partial T}\frac{\partial T}{\partial y}, \hspace{1cm} \text{at}~ z=H.
\end{equation}

The other boundary condition at the interface involves the vertical component of the velocity $v_z$ (non-permeability condition):

\begin{equation}
\label{eq:vertical-velocity}
v_z=0 \hspace{1cm}\text{at}~~ z=H.
\end{equation}

The thermal boundary condition at the free surface is:
\begin{equation}
 \label{eq:thermal-cond}
-\lambda\frac{\partial T}{\partial z}=h(T-T_0) \hspace{1cm} \text{at}~~z=H.
\end{equation}

The first term represents the heat conduction in the liquid, with $\lambda$ denoting the fluid thermal conductivity, while the second term expresses the heat flux density to the ambient air considered at $T_0$.

Finally, at the inlet, we use a half Poiseuille profile for velocity. This choice can be justified by considering a slip condition at the upper surface ($\frac{\partial v_x}{\partial z}=0, v_z=0$). The three-dimensional Poiseuille profile in the case of a rectangular channel with rigid walls has been computed analytically in \cite{Nicolas2000}. The inlet temperature is fixed at $T_0$:

\begin{equation}
\label{inlet-cond}
v_x=v_{\frac{1}{2}Pois(y,z)}; \hspace{0.5cm} v_y=v_z=0; \hspace{0.5cm} T=T_0 \hspace{0.5cm} \text{at}~~x=-2H.
\end{equation}

Using the channel height $H$, the mean flow velocity $U_m$ and $\rho U_m^2$ as reference quantities for the lengths, velocities and pressure, respectively, and using the reduced temperature $\theta=(T-T_0)/(T_c-T_0)$, the governing equations take the following dimensionless form: 

\begin{eqnarray}
 \vec{\nabla}.\vec{V}&=&0 \label{eq:ad-continuite}\\
(\vec{V}.\vec{\nabla})\vec{V}&=&-\vec{\nabla}
P+\frac{1}{Re}\vec{\nabla}^2\vec{V}+\frac{Ra}{PrRe^2}\theta\vec{k}\label{eq:ad-conservation}\\
\vec{V}.\vec{\nabla}\theta&=&\frac{1}{PrRe}\vec{\nabla}^2\theta \label{eq:ad-energie}
\end{eqnarray}

Thus, the dimensionless parameters that appear in these equations are the Reynolds number $Re=\frac{U_mH}{\nu}$, the Rayleigh number $Ra=\frac{g\beta(T_c-T_0)H^3}{\nu\alpha}$ and the Prandtl number $Pr=\frac{\nu}{\alpha}$.\\  
The boundary conditions (4) to (9) become:
\begin{eqnarray}
\vec{V}=\vec{0},\hspace{0.1cm} \frac{\partial \theta}{\partial Z}=0 ~~\text{for}~ X\in[-2,0[,\hspace{0.1cm} \theta=1
~~\text{for}~X\in[0,48] \hspace{0.4cm} \text{at}~ Z=0.\label{eq:ad-bottom-cond}\\
\vec{V}=\vec{0},\hspace{1cm} \frac{\partial \theta}{\partial Y}=0 \hspace{3.3cm} \text{at}\hspace{0.5cm} Y=0~~\text{and}~~Y=10.
\label{eq:ad-side-cond}\\
\frac{\partial V_X}{\partial Z}=-\frac{Ma}{Pe}\frac{\partial \theta}{\partial X},\hspace{0.5cm} \frac{\partial
V_Y}{\partial Z}=-\frac{Ma}{Pe}\frac{\partial \theta}{\partial Y}, \hspace{0.5cm} V_Z=0 \hspace{1cm} \text{at}~~ Z=1.
\label{eq:ad-shear-stress}\\
\frac{\partial \theta}{\partial Z}=-Bi\,\theta \hspace{7cm} \text{at}~~Z=1. \label{eq:ad-thermal-cond}\\
V_X=V_{\frac{1}{2}Pois(Y,Z)}, \hspace{0.5cm} V_Y=V_Z=0, \hspace{0.5cm} \theta=0 \hspace{1.5cm} \text{at}~~X=-2. \label{ad-inlet-cond}
\end{eqnarray}
Where $Ma=\frac{\partial\sigma}{\partial T}\frac{\Delta TH}{\mu\alpha}$, $Bi=\frac{hH}{\lambda}$ and $Pe=\frac{U_{m}H}{\alpha}$ are the Marangoni, Biot and Péclet numbers, respectively. To quantify the heat transfer at the bottom wall, we define the local, the average and the transverse average Nusselt numbers by: 
\begin{equation}\begin{split}
\label{eq:Nusselt}
Nu_l\left(X,Y\right) =-\biggl(\frac{\partial\theta}{\partial Z}\biggr)_{Z=0} &\hspace{1cm}\text{,} \hspace{1cm}
Nu_{av}=\frac{1}{S}\int\limits_S\!Nu_l\,\mathrm{d}S \\
Nu_{tav}\left(X\right) &=\frac{1}{10}\int_0^{10}\!Nu_l\,\mathrm{d}Y
\end{split}\end{equation}
with $S$ is the bottom heated surface of the channel.

\section{Numerical method}

\begin{figure}[h!]
  \begin{center}
\includegraphics[width=0.50\textwidth]{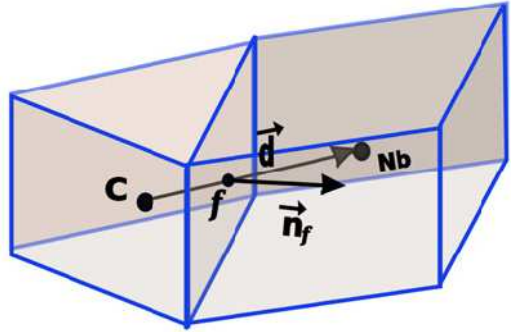} 
\caption{\it \small A computational cell C and one of its neighbors $N_b$.} \label{figure2}
  \end{center}
\end{figure}

The governing equations (\ref{eq:continuite}, \ref{eq:conservation} and \ref{eq:energie}) are solved using an in-house code called Tamaris. This code has a three-dimensional unstructured finite-volume framework that is applied to hybrid meshes. Variable values ($\vec{v}$, $p$ and $T$) are stored at cell centers in a collocated arrangement. Cell shapes can be of different forms (tetrahedral, hexahedral, prismatic or pyramidal). To describe the discretization method used in the code, we can write Eqs.(\ref{eq:conservation}) and (\ref{eq:energie}) in the generic convection-diffusion form with respect to a conserved variable $\phi$:

\begin{equation}
\label{eq:Num1}
\frac{\partial}{\partial
t}\int\limits_V\!\rho\phi\,\mathrm{d}V+\int\limits_S\!\rho\,\phi\vec{U}.\vec{n}\,\mathrm{d}S=\int\limits_S\!\Lambda\vec{\nabla}\phi.\vec{
n } \
, \mathrm{d}S+\int\limits_V\!S_{\phi}\,\mathrm{d}V
\end{equation}
Where $\Lambda$ is a diffusion coefficient and $S_{\phi}$ is a source term. The spatial schemes for approximating the diffusive and convective fluxes are both second-order accurate. The diffusion term is discretized by approximating the surface integrals with a sum over all cell faces $f$:
\begin{equation}
  \label{eq:Num2}
\int\limits_S\!\Lambda\vec{\nabla}\phi.\vec{n}\,\mathrm{d}S=\sum_{f}\Lambda_f A_f\bigl(\vec{\nabla}\phi\bigr)_f.\vec{n}_f
\end{equation}
where $A_f$ is the area of the face $f$. For unstructured meshes, orthogonality is an exception that needs to be handled correctly. Thus, the normal gradient $\bigl(\vec{\nabla}\phi\bigr)_f.\vec{n}_f$ is decomposed into an implicit contribution that uses the values of $\phi$ at the centers of the two cells sharing face $f$ (the first term on the RHS of Eq. (\ref{eq:Num3})) and a non-orthogonality correction term treated explicitly by a deferred approach to preserve the second-order accuracy of the centered differencing. We use the over-relaxed decomposition suggested by \cite{jasak1996} to enhance the convergence properties of the discretization of the diffusive term: 

\begin{equation}
  \label{eq:Num3}
\bigl(\vec{\nabla}\phi\bigr)_f.\vec{n}_f=\frac{\phi_{N_b}-\phi_c}{\lVert\vec{d}\rVert}\frac{1}{\vec{d}.\vec{n}_f}+\overline{\vec{
\nabla}\phi } .\left( \vec{n}_f-\frac{\vec{d}}{\vec{d}.\vec{n}_f}\right) 
\end{equation}
$\vec{d}\left(d_x,d_y,d_z\right)$ is the vector joining the centers of the two adjacent cells (see Figure \ref{figure2}). The average gradient $\overline{\vec{ \nabla}\phi}$ is interpolated from the gradients of these neighboring cells. The gradients of the variables at the cell centers are computed by a least-squares method where, for each computational cell, the following $3\times3$ linear system is solved:
\begin{equation}
  \label{eq:Num4}
\mathbf M\cdot\vec{\nabla}\phi=\vec{H}
\end{equation}
where the components of the matrix $\mathbf{M}$ and the vector $\vec H$ are written as:
\[ M_{ij}=\sum_{k=1}^{N_v}\frac{1}{\lVert\vec{d_k}\lVert^2}d_{k,j}d_{k,j} \hspace{0.1cm}\text{and}\hspace{0.1cm}
H_i=\sum_{k=1}^{N_v}\frac{1}{\lVert\vec{d_k}\lVert^2}d_{k,j}\left(\phi_k-\phi_C\right)\]\\  
where $N_v$ is the number of surrounding cells. In our case, we choose to include all of the cells that share a vertex with cell $C$.
Thus, in the case of a mesh formed by structured hexahedral cells, $N_v$ can reach 26. The contribution of each cell is weighted by a factor equal to the inverse of the square of the distance between the two cell centers $\frac{1}{\lVert\vec{d_k}\lVert^2}$.\\  
The convection terms in Eq.\ref{eq:Num1} are also transformed into a sum over the cell faces $f$ by decomposing the surface S:
\begin{equation}
  \label{eq:Num7}
\int\limits_S\!\rho\phi\vec{U}.\vec{n}\,\mathrm{d}S=\sum_{f}(\rho\phi A)_f \vec{U}_f. \vec{n}_f
\end{equation}
where the face values $\phi_f$ require appropriate interpolation to be accurate and bounded. Thus, we use the non-linear
high-resolution (HR) bounded scheme CUBISTA by \cite{Alves} in the $\Upsilon$ formulation of \cite{Ng2007},
where they expressed $\phi_f$ as a function of the upwind (UP) value of $\phi$ and its centered differencing (CD) value:
\begin{equation}
  \label{eq:Num8}
\phi_f^{HR}=\phi^{UP}+\Upsilon\Bigl(\phi_f^{CD}-\phi^{UP}\Bigr)
\end{equation}
The coefficient $\Upsilon$ is determined for each face based on the local shape of the flow solution using the normalized variable diagram (NVD) framework and observing the convection boundedness criterion (CBC) \citep{Gaskell}. The first term of the RHS of Eq. (\ref{eq:Num8}) is accounted for implicitly, while the second term is treated explicitly with the deferred-correction practice \citep{ferziger1999}.
The pressure-velocity coupling is ensured by the SIMPLE algorithm \citep{patankar1980}, whereas the mass fluxes at the cell faces are evaluated using the Rhie-Chow interpolation \citep{RhiCho83} to avoid checker-boarding in the pressure field. Within each iteration of the SIMPLE algorithm, the energy equation is solved after the resolution of the momentum equation and the Poisson equation for the pressure correction, and then the next SIMPLE iteration starts unless the convergence for $\vec{v}$, $p$, and $T$ is achieved. The resolution of the energy equation is integrated in the SIMPLE iteration to account for the high level of its coupling with the momentum equation through the body forces term and with the thermocapillary boundary conditions. This code is parallelized following a non-overlapping domain decomposition approach and using the Petsc \citep{petsc} library.

\section{Code validation}
\begin{figure}[h!]
  \begin{center}
\includegraphics[width=\linewidth]{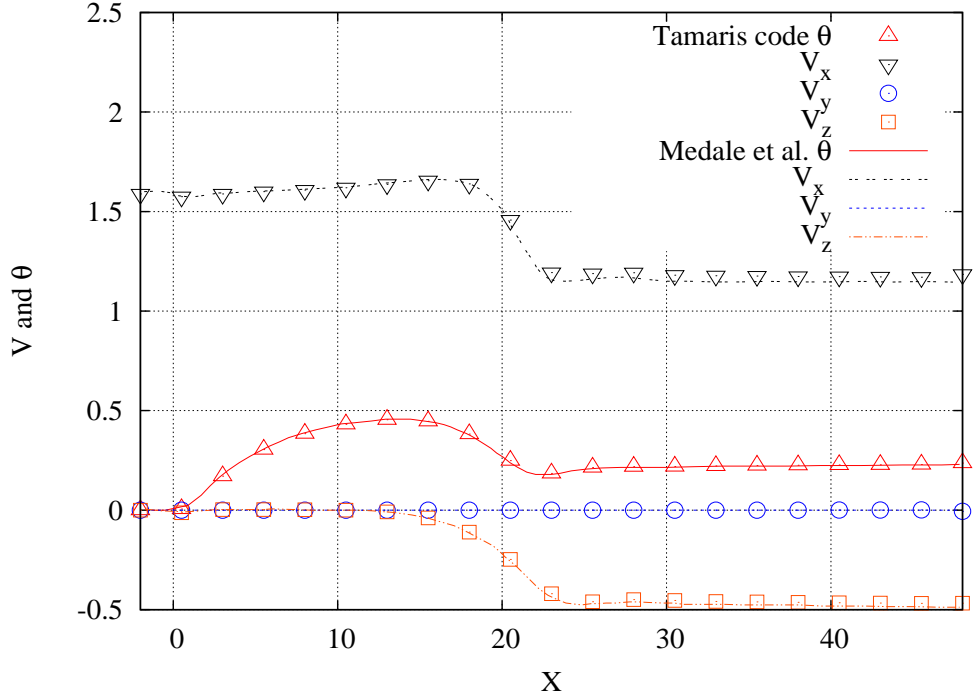}
\caption{\it \small The longitudinal profiles of the rescaled variables $V_X$, $V_X$, $V_Z$ and $\theta$ (represented by symbols) along the axis ($Y=5$, $Z=0.5$) compared with the reference solution \citep{Necolas-Bench} (lines).}\label{validation}
\end{center}
\end{figure}

\begin{figure}[h!]
  \begin{center}
\includegraphics[width=\linewidth]{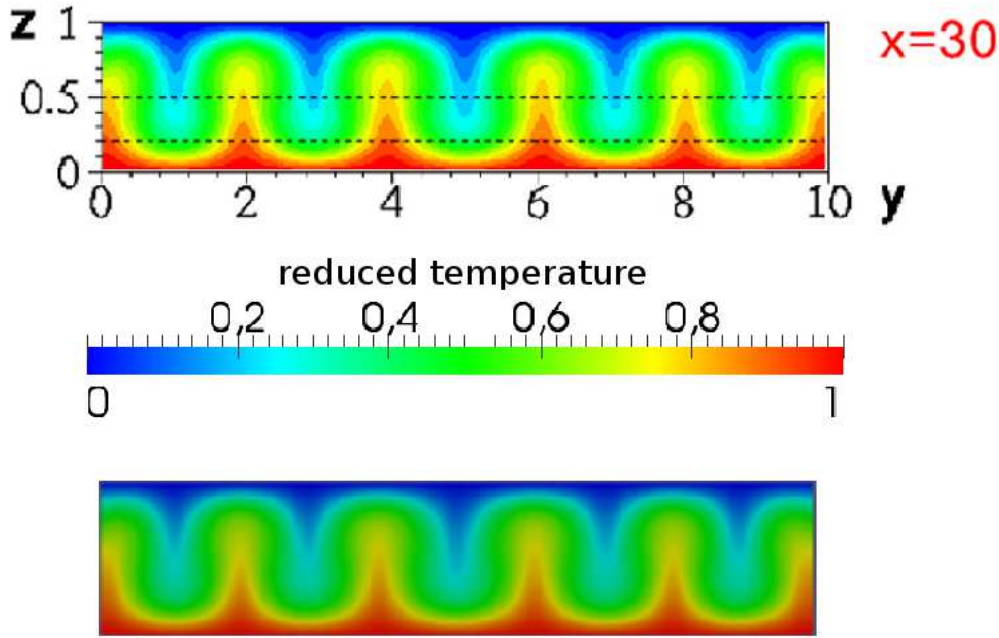}
\caption{\it \small The reduced temperature field in the vertical plane ($X=30$). The reference solution \citep{Necolas-Bench} (top) and the
Tamaris code results.
(bottom).}\label{Tfield}
\end{center}
\end{figure}

\begin{figure}[h!]
  \begin{center}
\includegraphics[width=0.60\textwidth]{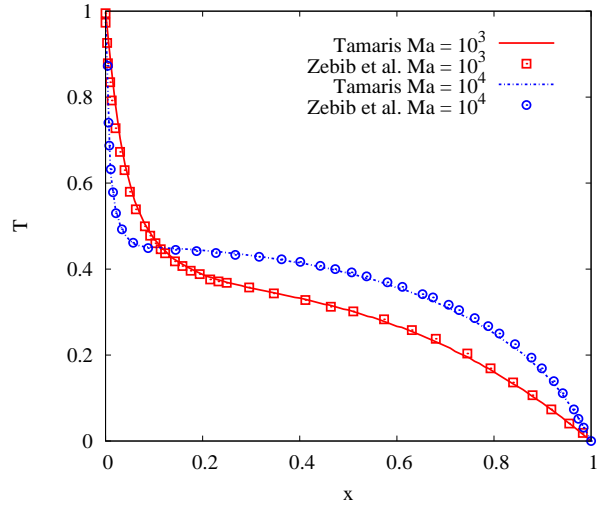} 
\caption{\it \small Comparison of the temperature profile obtained on the free surface of a two dimensional square cavity for
Marangoni number values of $Ma=10^3$ and $Ma=10^4$ with the results of \cite{zebib} ($Pr=1$).}\label{MarCav}
  \end{center}
\end{figure}

This code has been successfully validated in some situations involving flow and heat transfer \citep{elomari2010b,elomari2011a}. For this study, we conducted additional code validations. The first supplementary validation concerns the flow in a horizontal channel heated uniformly from below, which was first presented by \cite{Necolas-Bench} (steady Poiseuille-Rayleigh-Bénard flow). Quantitative comparisons have been performed using a non-uniform mesh resolution $n_x=128$, $n_y=176$ and $n_z=50$. Figure \ref{validation} shows the longitudinal profiles of the rescaled components of the velocity and reduced temperature along the axis for the coordinates $Y=5$ and $Z=0.5$. Figure \ref{Tfield} shows the reduced temperature field in the vertical plane of $X=30$. As observed in Figures \ref{validation} and \ref{Tfield}, the present calculations and those of Nicolas et al. agree well.\\
The second validation test case concerns the Marangoni convection. It is validated for different parameter combinations for a flow in a 2D rectangular cavity with a free surface. The aspect ratio is fixed at $\Gamma=1$, and the side walls are maintained at a constant temperature difference $\Delta T$. The control parameters of the validation test case are as follows: the thermocapillary Reynolds number, defined as: $Re_{th}=\frac{\gamma\Delta T H}{\rho\nu^2}$, the Prandtl number $Pr$, (the Marangoni number being $Ma=Re_{th}Pr$) and the dynamical Bond number $Bd=\frac{\rho g\beta H^2}{\gamma}$, which characterizes the gravity level. This last parameter has the advantage of immediately giving the importance of the buoyancy force relative to the thermocapillary force. We perform different numerical simulations for the same conditions as those described in Table \ref{tab:valid} using a uniform 121x121 grid. In Figure \ref{MarCav}, we plot the temperature profile obtained for $Ma=10^3$ and $Ma=10^4$. The results are compared with those given in Ref. \citep{zebib} and exhibit a satisfactory correspondence. Other comparisons with data previously published by \citep{carpenter1990,kuhlmann,XuandZebib,carpenter1989} are made; these results are presented in Table \ref{tab:valid}. Our calculations are in good agreement with the data from the literature.

\begin{table}[ht!]
\begin{center}
{\footnotesize
\begin{tabular}{|c|c|c|c|c|c|c|}\hline
Parameters& Author/Grid & u(0,0.5)& $Nu_{x=0.5}$ & $Nu_{x=-0.5}$ \\
\hline
                & \cite{XuandZebib}     & -305    & 4.36  & 4.36  \\
$Re_{th}=10000$ & \cite{carpenter1990}  & -296    & 4.33  & 4.36  \\
$Pr=1$          & \cite{kuhlmann}       & -304.36 & 4.364 & 4.364 \\
$Ma=10000$      & (115x115)             &         &       &       \\
$Bd=0$          & \cite{kuhlmann}       & -304.27 & 4.363 & 4.363 \\
                &     (211x211)         &         &       &       \\
                & Tamaris (121x121)     & -301.99 & 4.358 & 4.361 \\
\hline
$Re_{th}=5000$ & \cite{carpenter1989}   & -179    & 4.17  & 4.14  \\
$Pr=1$         & \cite{kuhlmann}        & -175.90 & 4.546 & 4.545 \\
$Ma=5000$      &  (115x115)             &         &       &       \\
$Bd=10$        & \cite{kuhlmann}        & -175.97 & 4.545 & 4.545 \\
               &  (211x211)             &         &       &       \\
               & Tamaris (121x121)      & -174.38 & 4.543 & 4.543 \\
\hline
               & \cite{XuandZebib}      & -29.8   & 6.61  & 6.61  \\
$Re_{th}=2000$ & \cite{carpenter1990}   & -37.2   & 6.60  & 6.42  \\
$Pr=30$        & \cite{kuhlmann}        & -24.22  & 6.26  & 6.26  \\
$Ma=60000$     & (115x115)              &         &       &       \\
$Bd=0$         & \cite{kuhlmann}        & -23.98  & 6.234 & 6.234 \\
               & (211x211)              &         &       &       \\
               & Tamaris (121x121)      & -24.64  & 6.135 & 6.235 \\
\hline
\end{tabular}
}
\caption{\it \small The results for the validation test case corresponding to thermocapillary flow in a two-dimensional open square cavity compared with the results from the literature \citep{carpenter1990,kuhlmann,XuandZebib,carpenter1989}. u(0,0.5) is the velocity at the center of the free surface. $Nu_{x=0.5}$ and $Nu_{x=-0.5}$ are the mean Nusselt numbers over the two vertical walls of the cavity.}
\label{tab:valid}
\end{center}
\end{table}

\section{Results and discussion}
To study the influence of thermal instabilities on the patterns of the flow and heat transfer characteristics in a horizontal channel with a free surface and uniformly heated from below, we examine several numerical results for different dimensionless parameters that control the physics of the problem. These parameters are the Reynolds number, the Rayleigh number, the Biot number and the Marangoni number. The numerical results are obtained for a Prandtl number fixed at $Pr=7$. We are interested in using this work to determine the thermal instabilities that exhibit stationary longitudinal rolls when the Rayleigh and the Reynolds numbers are between $2500\leq Ra\leq 22500$ and $5\leq Re\leq 20$, respectively. After a grid size-dependence study (see Appendix A), a mesh of resolution $n_x=128$, $n_y=176$ and $n_z=50$ was chosen for this study. 
\subsection{Reynolds number effect}
\begin{figure}[h!]
  \begin{center}
\includegraphics[height=7cm , width=0.80\textwidth]{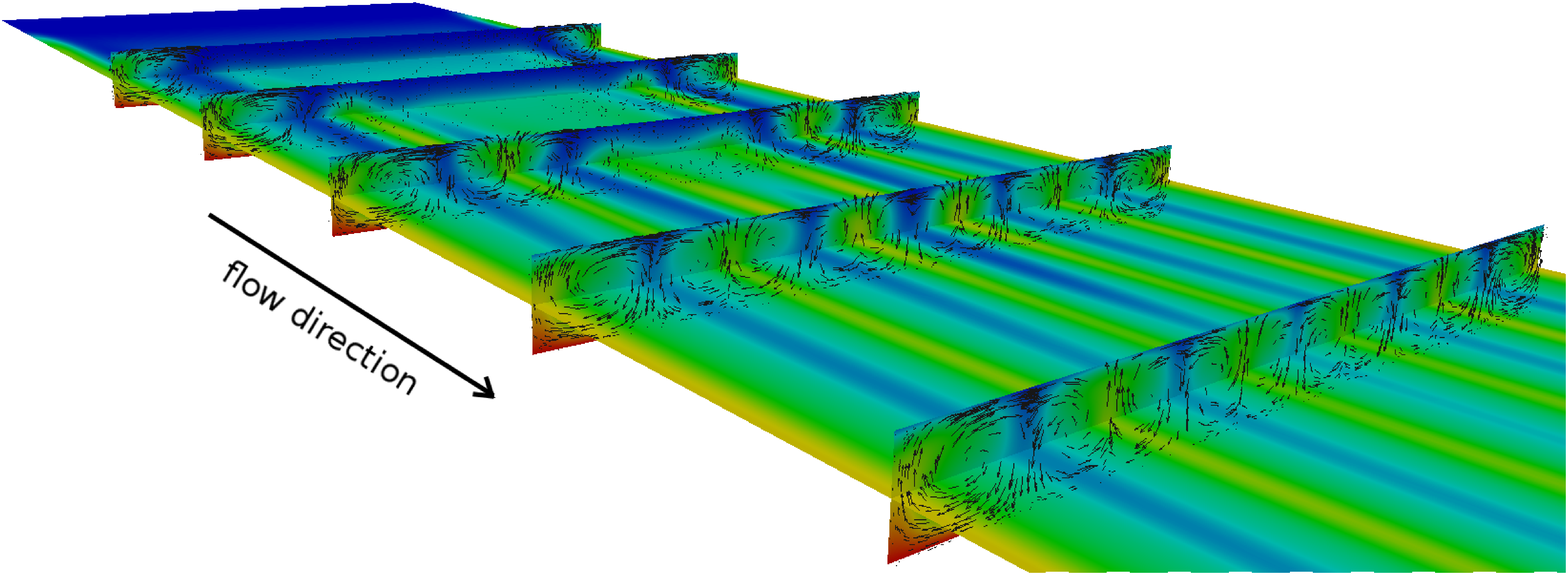}
\caption{\it \small Visualization of the longitudinal rolls at selected cross sections ($Re=15$, $Ra=5000$, $Bi=15$ and
$Ma=500$.}\label{flow}
  \end{center}
\end{figure}

\begin{figure}[h!]
  \begin{center}
\includegraphics[width=\linewidth]{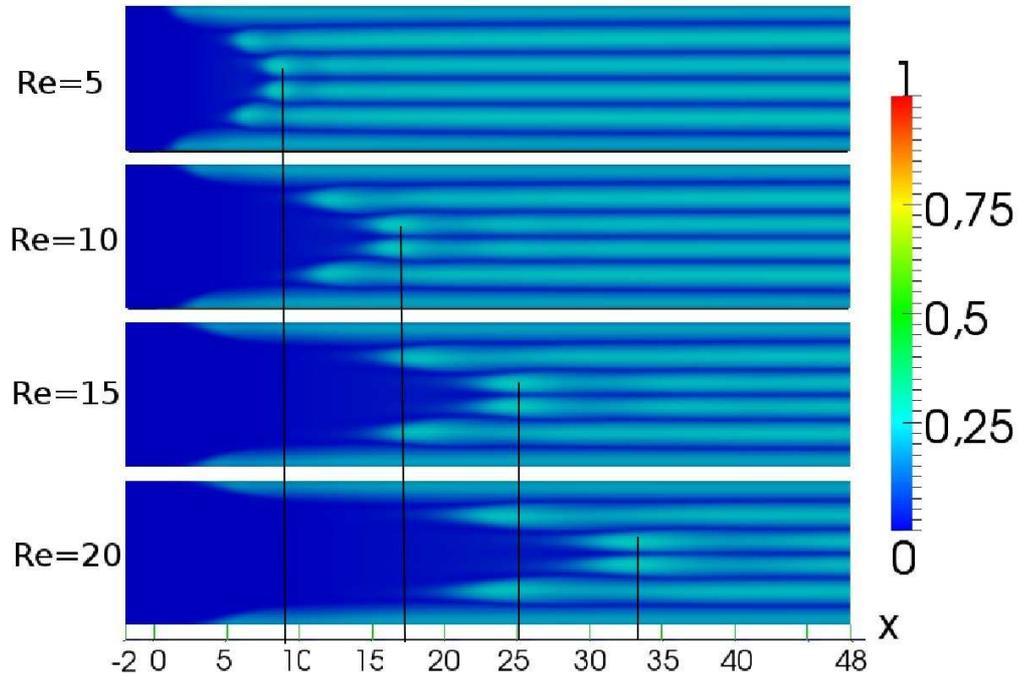}
\caption{\it \small The reduced temperature field $\theta$ on the free surface for different Reynolds numbers $Re$ ($Bi=15$, $Ra=5000$,
$Ma=500$).}\label{Re-effect}
\end{center}
\end{figure}

\begin{figure}[h!]
  \begin{center}
\includegraphics[width=\linewidth]{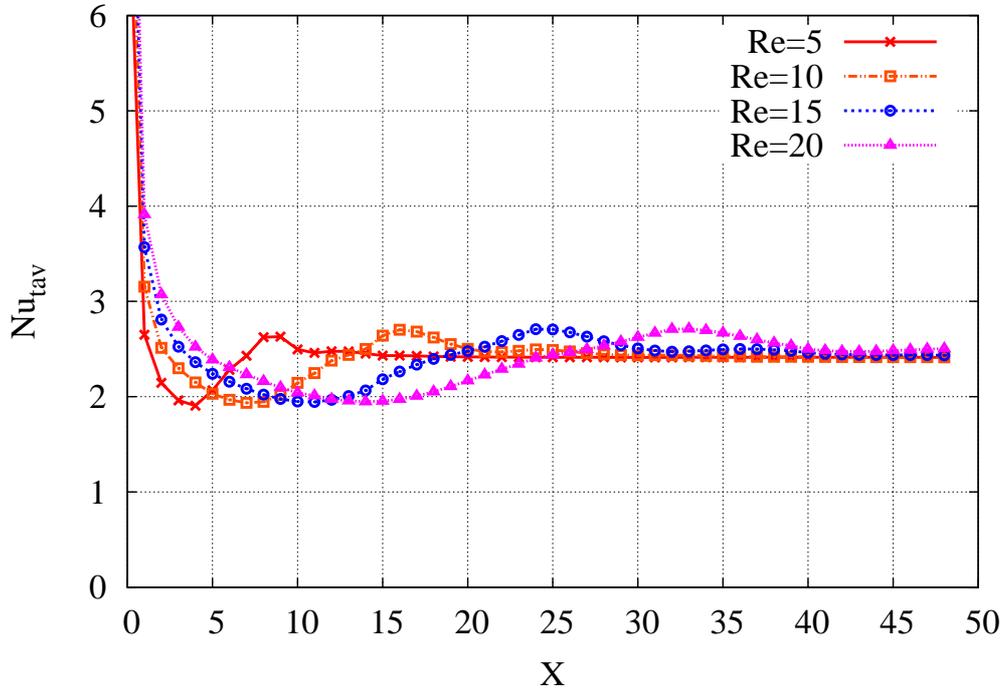}
\caption{\it \small The longitudinal variation of the transversally averaged Nusselt number $Nu_{tav}$ for different Reynolds numbers
$Re$ ($Bi=15$, $Ra=5000$, $Ma=500$).}\label{Nu-Re.eps}
\end{center}
\end{figure}

As illustrated in Figure \ref{flow}, the longitudinal rolls are first initiated near the side walls and grow gradually toward the center of the channel in the downstream direction. Indeed, the predominant heat transfer mechanism in these zones is the thermal conduction leading to a local temperature increase because the forced convection is weak in the vicinity of the lateral walls. This fluid heating induces natural convection currents near these lateral walls, causing the onset of the first longitudinal rolls. This onset happens near the flow inlet. The transverse momentum transport causes the progressive formation of other rolls toward the center of the channel as the fluid progress downstream. The rolls' dimension is nearly the same as the channel height, and the fully developed section contains 10 rolls. To highlight the Reynolds number's effect on the development of longitudinal rolls, we present the temperature field at the free surface for different values of the Reynolds number $(5\leq Re\leq 20$ and for Biot, Marangoni and Rayleigh numbers, respectively set to: $Bi=15$, $Ra=5000$ and $Ma=500$ in Figure \ref{Re-effect} (this value of $Ma$ corresponds to weak thermocapillary effects, as will be shown in section 5.4). These flows correspond to a Richardson number ranging from 1 to 29. We found that an increased Reynolds number causes the downstream displacement of the onset position of the longitudinal rolls. This effect is particularly true for central rolls, which appear at a position $X=8$ for $Re=5$, while they appear at a position $X=32$ for $Re=20$. On the other hand, the rolls near the side walls form almost at the same position, regardless of the Reynolds number.\\
Figure \ref{Nu-Re.eps} shows the variation along the channel of the transversally averaged Nusselt number $Nu_{tav}$ (cf. Eq. \ref{eq:Nusselt}) for different Reynolds numbers to evaluate the influence of the Reynolds number on heat transfer. We note that $Nu_{tav}$ first decreases and then increases to reach a maximum value before decreasing again slightly (behavioral characteristic of mixed convection). The maximum $Nu_{tav}$ positions correspond to the positions of the appearance of the central rolls. We also observe that the number of rolls (10) is not affected by Re in the studied range. At low Reynolds numbers, the natural convection improves the heat transfer for a short distance after the entry channel, causing $Nu_{tav}$ to peak near $X=8$. This peak has almost the same value for all Re numbers, but at different positions. Table \ref{tab:av-Nu} presents the evolution of the average Nusselt number $Nu_{av}$ over the whole heated wall with respect to the Reynolds number $Re$. We note that $Nu_{av}$ increases slightly with increasing $Re$, which can be explained by the transition from a mixed convection regime, where the heat transfer is ensured by buoyancy currents, toward a rather forced convection regime (relative importance of natural convection becomes negligible for $Ri<0.1$), where the heat transfer is provided mainly by the forced convective flow.

\begin{table}[h!]
 \begin{center}
  \begin{tabular}{|c|c|c|c|c|}\hline
Re & 5 & 10 & 15 & 20 \\
\hline
Ri & 28.57 & 7.14 & 3.17 & 1.78 \\
\hline
$Nu_{av}$ & 2.31 & 2.35 & 2.39 & 2.42 \\
\hline
 \end{tabular}
\caption{\it \small The evolution of the average Nusselt number $Nu_{av}$ for different Re ($Bi=15$, $Ra=5000$, $Ma=500$).}
\label{tab:av-Nu}
\end{center}
\end{table}

\subsection{Rayleigh number effect}

\begin{figure}[h!]
  \begin{center}
\includegraphics[height=10cm ,width=0.80\linewidth]{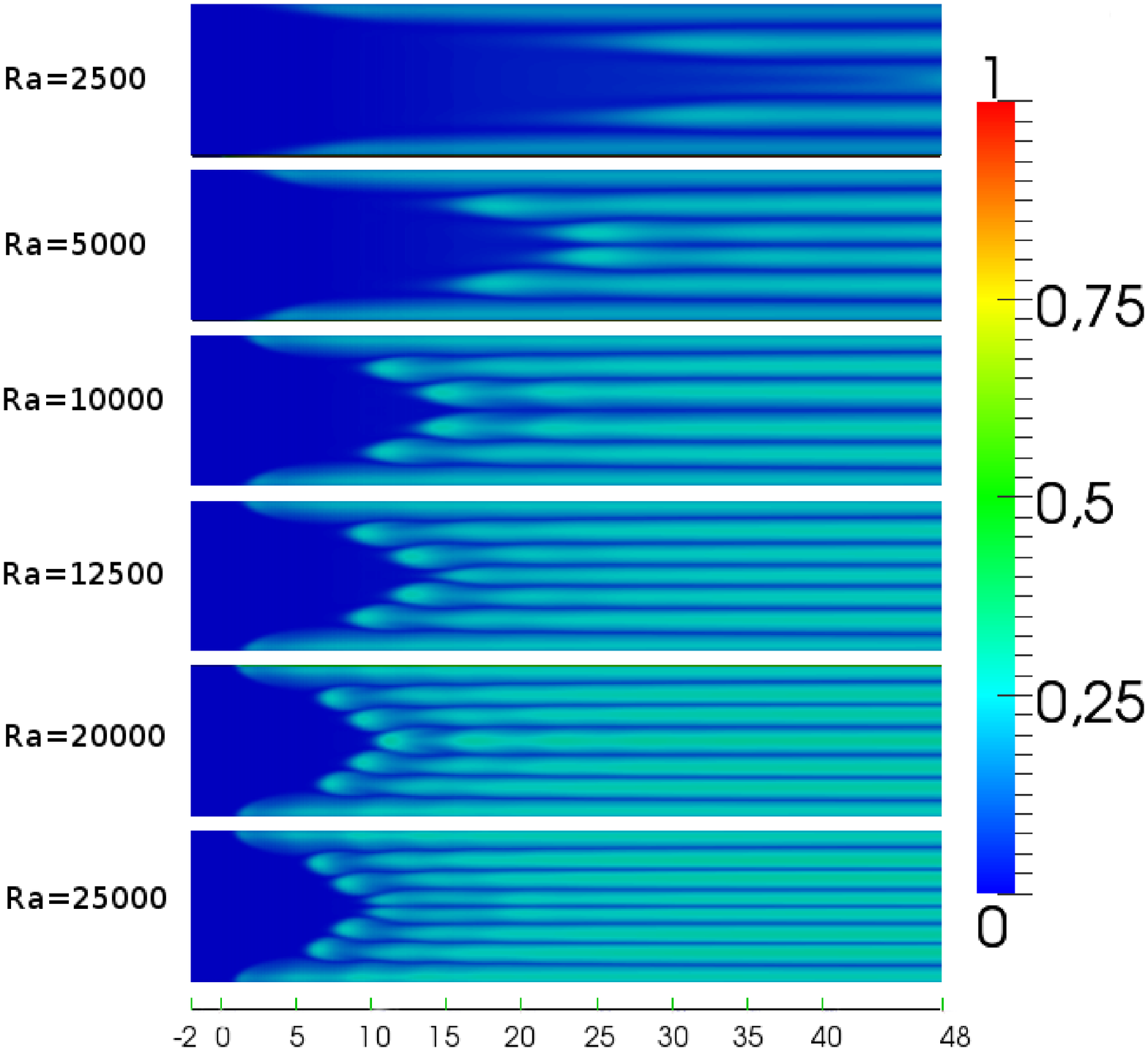}
\caption{\it \small The reduced temperature fields $\theta$ on the free surface for different Rayleigh numbers $Ra$ ($Bi=15$, $Re=15$,
$Ma=500$).
}\label{effet-Ra}
\end{center}
\end{figure}

\begin{figure}[h!]
  \begin{center}
\includegraphics[width=\linewidth]{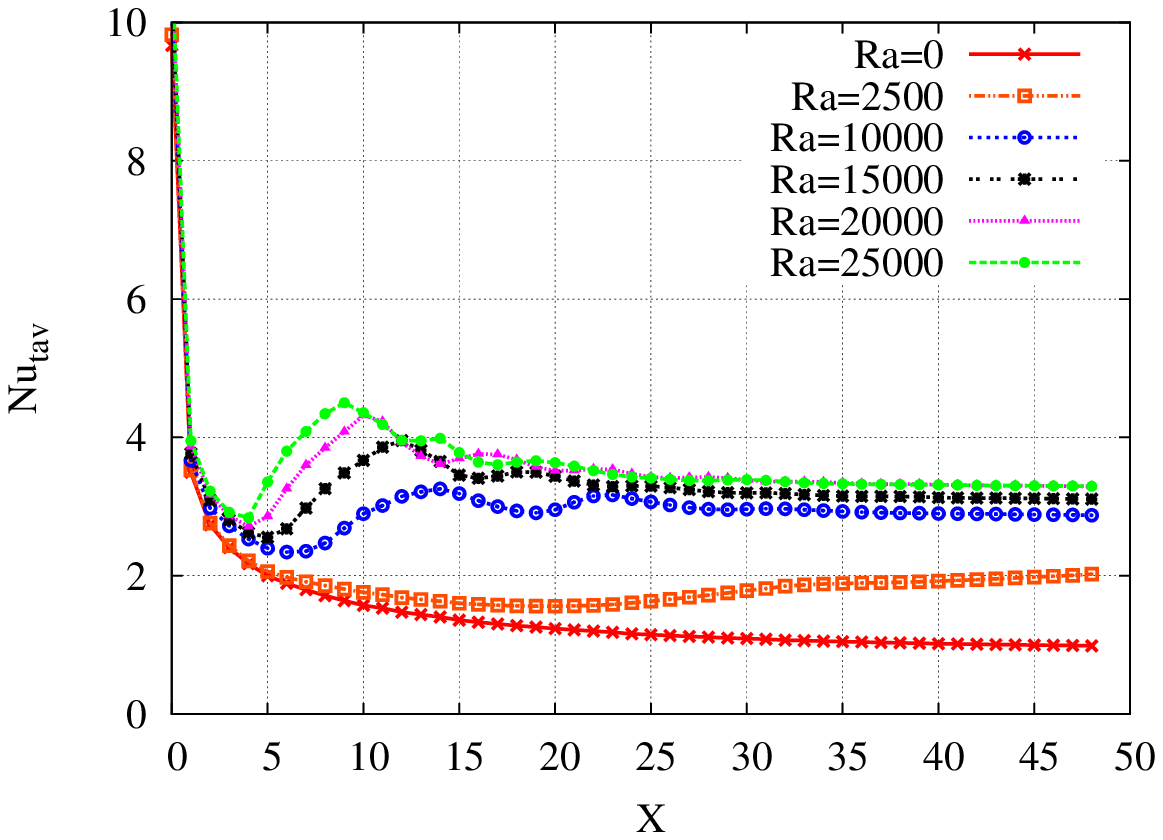}
\caption{\it \small The longitudinal variation of the transversally averaged Nusselt number $Nu_{tav}$ for different Rayleigh numbers $Ra$
($Bi=15$, $Re=15$, $Ma=500$).}\label{Nu-Ra}
\end{center}
\end{figure}

Figure \ref{effet-Ra} illustrates the effect on the temperature field at the free surface for different values of the Rayleigh number $Ra$. The other flow parameters are fixed: $Bi=15$, $Re=15$ and $Ma=500$. One can observe that the increase in the Rayleigh number shifts the onset position of the longitudinal rolls towards the entrance of the canal. This behavior is especially true for the central rolls, while the rolls near the side walls are formed at effectively the same X position (as in the previous section, the downstream displacement of the rolls is related to a decrease in the Richardson number). The morphology of the flow changes when $Ra$ increases; in fact, the number of rolls increases, from 10 rolls for relatively low Ra values to 12 rolls for $Ra\geq12500$. Figure \ref{Nu-Ra} shows the longitudinal variation of $Nu_{tav}$ with respect to the Rayleigh number. We note that $Nu_{tav}$ increases with increasing $Ra$ due to the onset of natural convection, which appears near the entrance for large Rayleigh numbers ($Ra\geq12500$), and to the increase in the kinetic energy generated by the existence of more longitudinal rolls. The profiles of $Nu_{tav}$ also indicate the position where the rolls are completely developed in the cross section, which corresponds to the maximum Nusselt number. The slight oscillations observed in the $Nu_{tav}$ evolutions are associated with longitudinal modulations in the shape of the rolls where they start to develop. For $Ra=0$, the case for which we do not observe roll development, the Nusselt number decreases continuously from the beginning of the heating zone to the channel outlet.

\subsection{Biot number influence}

\begin{figure}[h!]
  \begin{center}
\includegraphics[width=\linewidth]{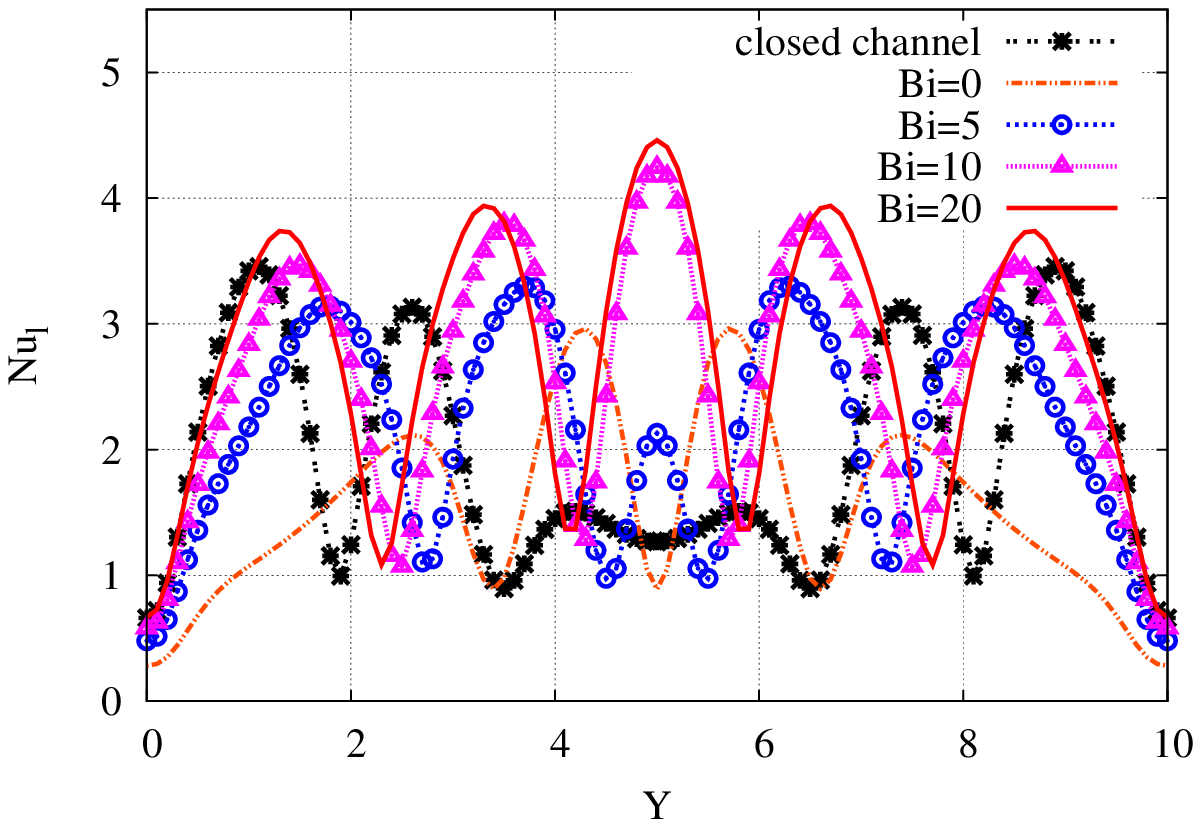}
\caption{\it \small The transverse variation of the local Nusselt number at $X=25$ for different Biot numbers ($Re=15$, $Ra=5000$,
$Ma=500$).
}\label{Nu-Bix25}
\end{center}
\end{figure}

\begin{figure}[h!]
  \begin{center}
\includegraphics[width=\linewidth]{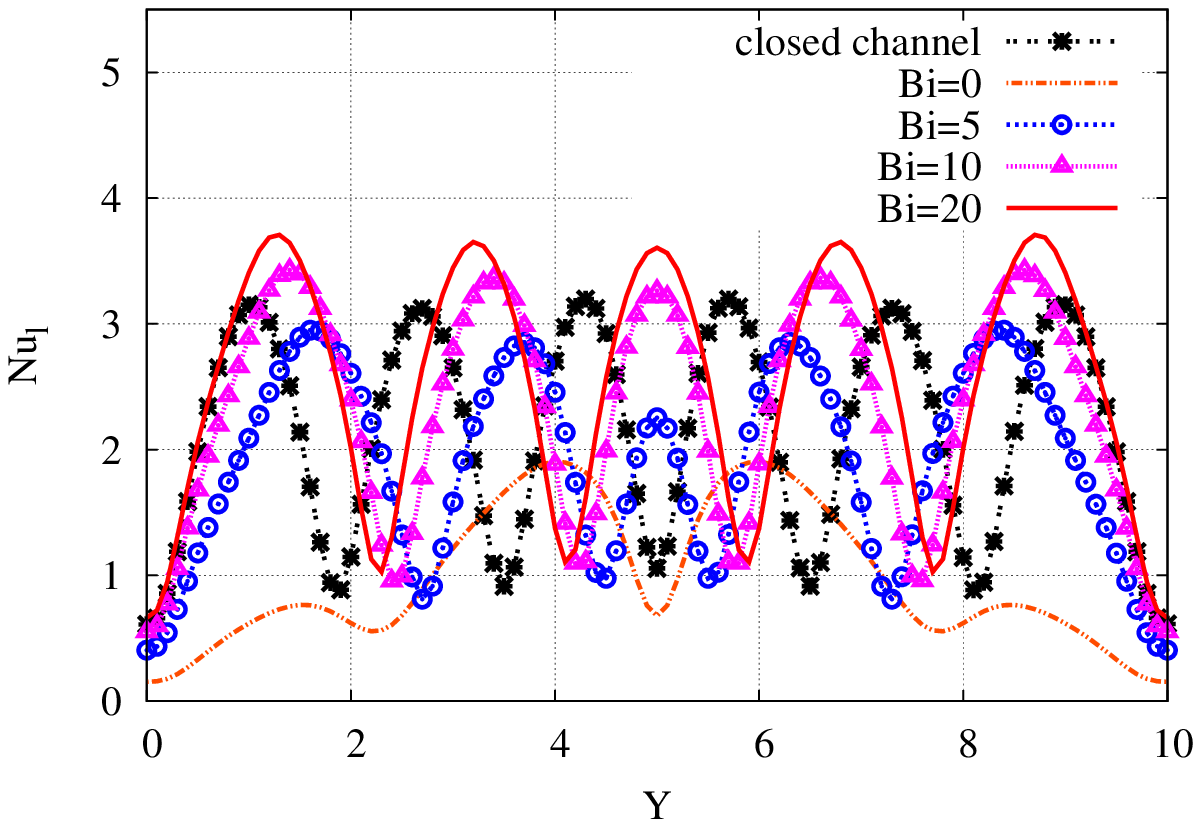}
\caption{\it \small The transverse variation of the local Nusselt number at $X=40$ for different Biot numbers ($Re=15$, $Ra=5000$,
$Ma=500$).}\label{Nu-Bix40}
\end{center}
\end{figure}

\begin{figure}[h!]
  \begin{center}
\includegraphics[width=\textwidth]{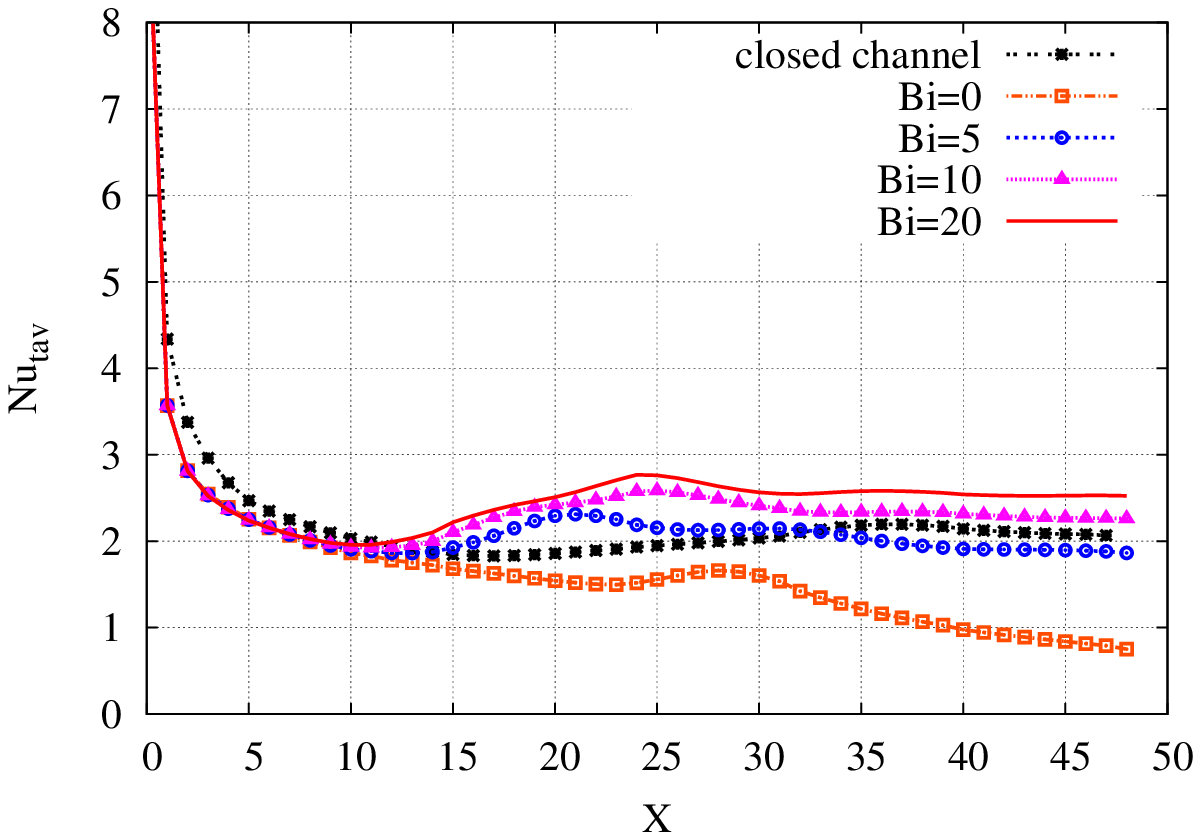} 
\caption{\it \small The longitudinal variation of the transverse average Nusselt number for different Biot numbers ($Re=15$, $Ra=5000$ and
$Ma=500$).}\label{nu_tav}
  \end{center}
\end{figure}

In Figures \ref{Nu-Bix25} and \ref{Nu-Bix40}, we present the transverse variation of the local Nusselt number at the bottom wall for various Biot numbers at $X=25$ (beginning of the rolls' development in the cross-section) and $X=40$ (all of the rolls of the cross-section are completely developed), respectively. The Biot number measures the ratio of the internal thermal resistance of the liquid film to the thermal resistance between the free surface and the surrounding air; a null Biot number corresponds to an adiabatic free surface. The other flow parameters are fixed at $Re=15$, $Ra=5000$ and $Ma=500$. We observe in these figures that the parietal heat transfer has a sinusoidal distribution for all of the studied cases. The distribution and number of peaks are related to the velocity distribution in the fluid generated by the ascent of hot fluid plumes under the action of buoyancy forces. They, therefore, reflect the organization of thermal instabilities into longitudinal rolls. In comparison, the results for a closed channel are also given (PRB flow). At $X=25$ (Figure \ref{Nu-Bix25}), the Nusselt number evolution of the closed channel has six peaks. The two central peaks are of very low amplitude due to the beginning of the center rolls development. This is not the case when considering solutions for the free surface, where we see four peaks for $Bi=0$ and five peaks for $Bi\geq5$ and whose amplitude is higher in the center and grows with an increasing Biot number. When the heat exchange on the upper surface is very low $(Bi\backsimeq0)$, the liquid film temperature increases in a generalized way, which reduces the thermal gradients and noticeably reduces the heat transfer at the bottom wall. Indeed, the convective roll number is lower for $Bi=0$ compared to other values. The hydrodynamic field and the $Nu_l$ distribution are related; indeed, every peak of the $Nu_l$ curve corresponds to a pair of rolls. Thus, 8 rolls are formed for $Bi=0$ and 10 rolls form for $Bi\geq5$. For $Bi=5$, the two central rolls have a lower intensity than the side rolls, regardless of the $X$ position, which is opposite to the behavior for $Bi=10$ or $Bi=20$, which present similar distributions with regular peaks from $X=40$, but with slightly greater $Nu_l$ values for $Bi=20$ (Figure \ref{Nu-Bix40}). The presence of an upper wall (case of closed channel) increases the number of rolls to 12. The imposed temperature $T_0$ at the top wall increases the thermal gradient in the film and should tend to improve the heat exchange. However, this favorable effect is counterbalanced by the viscous effects at this wall that decrease the fluid velocity. Thus, the Nusselt number values for the closed channel are lower than those obtained for $Bi=10$ and $Bi=20$. We also observe that the rolls in the case of a closed channel are regular (same wavelength), while in the case of the channel with free surface, the rolls close to the walls have a wavelength greater than those in the center.\\
Figure \ref{nu_tav} shows the longitudinal variation of the transversally average Nusselt number $Nu_{tav}$ for different Biot numbers. For the channel with the free surface, the $Nu_{tav}$ decreases first until $X=10$ and then increases gradually to a maximum located between $X=22$ and $X=28$, which corresponds to a position where all of the rolls are developed in the cross section. For the closed channel, the maximum is axially reached further downstream, at $X=37$. Indeed, the presence of the upper rigid wall stabilizes the flow and delays the development of longitudinal rolls.

\subsection{Marangoni effect}
\begin{figure}[h!]
  \begin{center}
\includegraphics[height=7cm , width=0.90\textwidth]{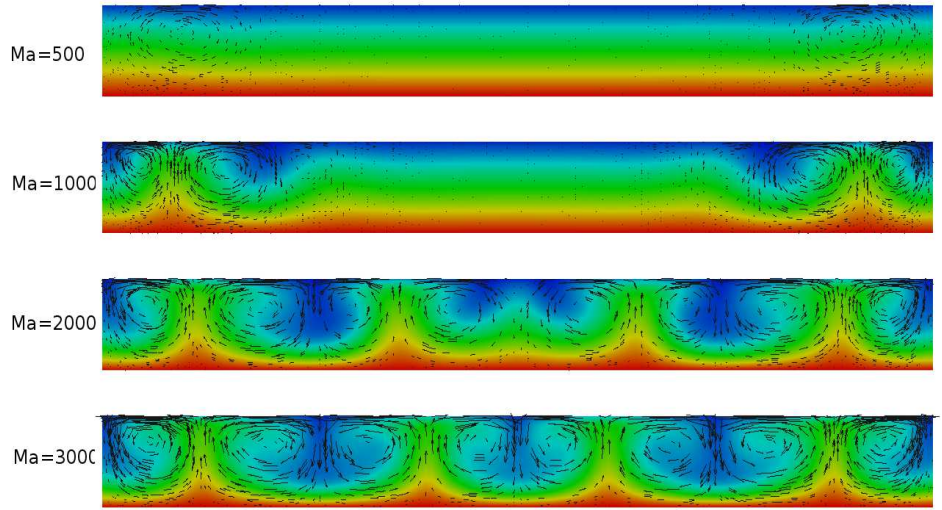} 
\caption{\it \small The longitudinal vortex rolls due to Marangoni effect at cross section $X=40$.}\label{conv-ma}
  \end{center}
\end{figure}

\begin{figure}[h!]
  \begin{center}
\includegraphics[width=\linewidth]{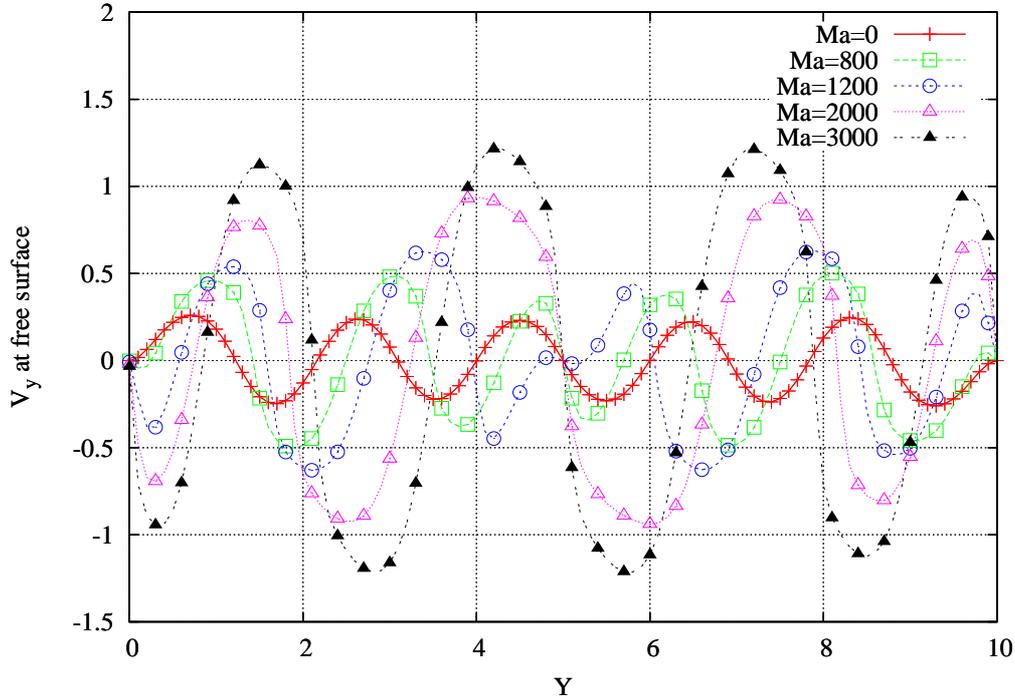}
\caption{\it \small The transverse variation at $X=30$ and $Z=1$ of the velocity $V_y$ for different Marangoni numbers $Ma$ ($Re=15$,
$Ra=5000$, $Bi=15$). }\label{vy-ma}
\end{center}
\end{figure}

\begin{figure}[h!]
  \begin{center}
\includegraphics[width=\linewidth]{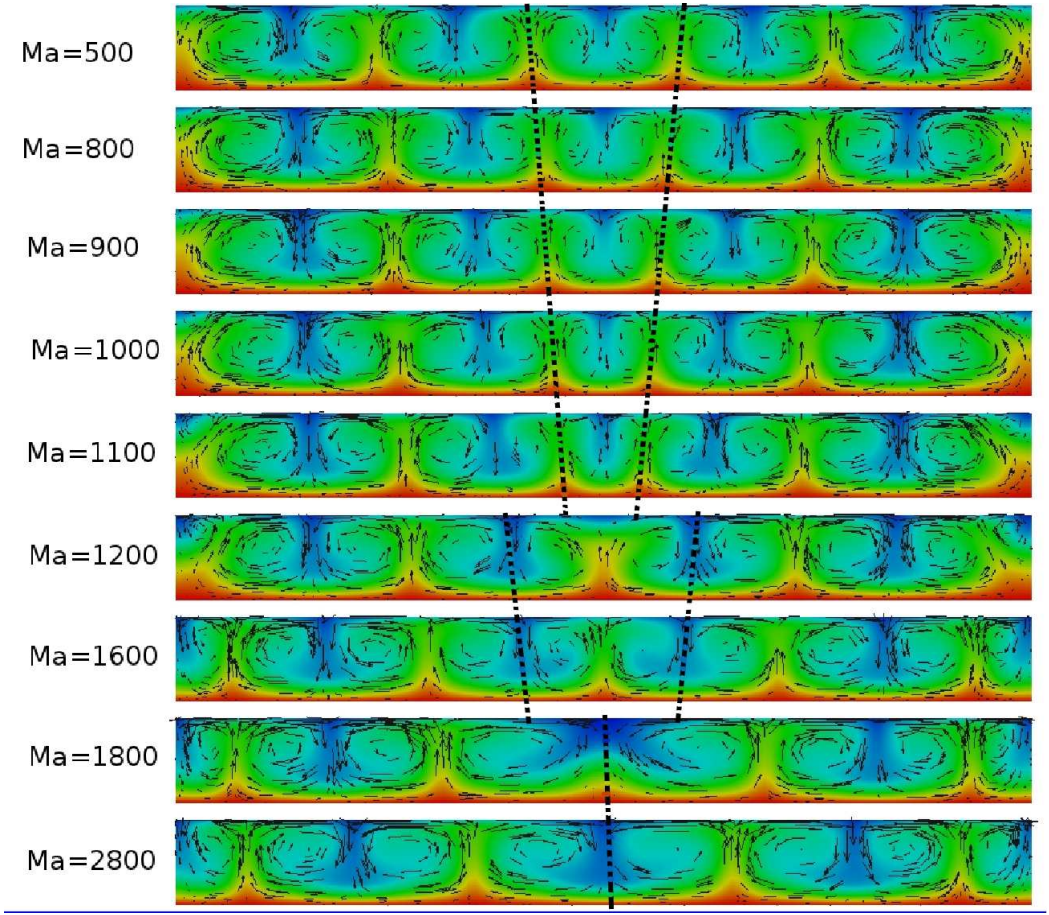}
\caption{\it \small The vertical cross-section at $x=30$ for different Marangoni numbers $Ma$ ($Ra=5000$, $Re=15$ and
$Bi=15$).}\label{effet-ma}
\end{center}
\end{figure}

\begin{figure}[h!]
  \begin{center}
\includegraphics[width=\linewidth]{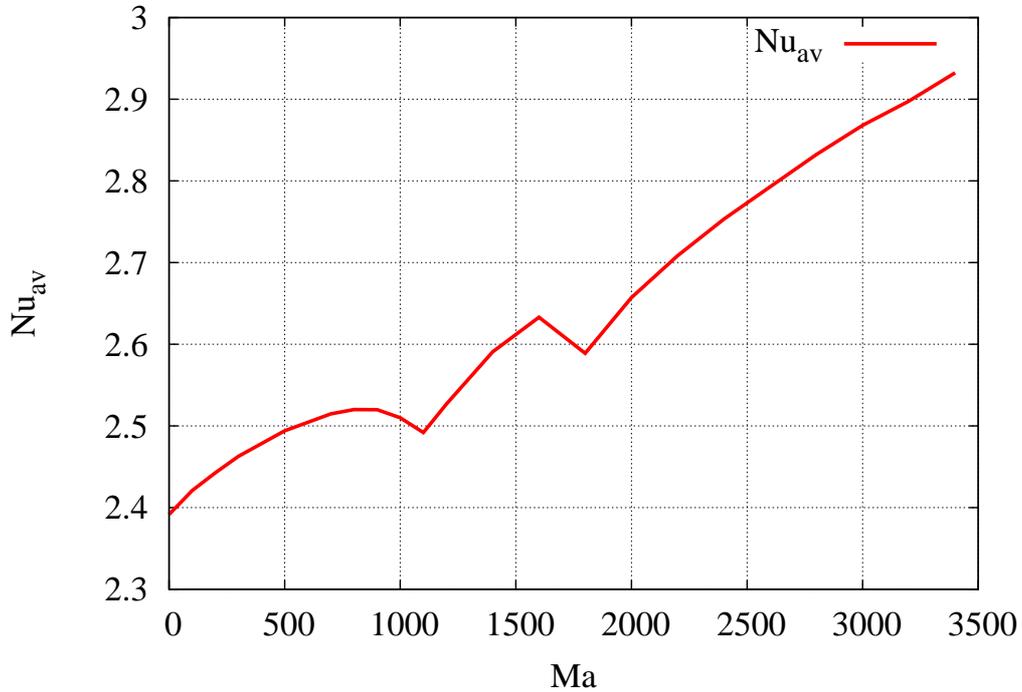}
\caption{\it \small The variation of the average Nusselt number at the heated wall with respect to the Marangoni number $Ma$ ($Ra=5000$,
$Re=15$ and $Bi=15$).}\label{nu-ma}
\end{center}
\end{figure}

\begin{figure}[h!]
  \begin{center}
\includegraphics[width=\linewidth]{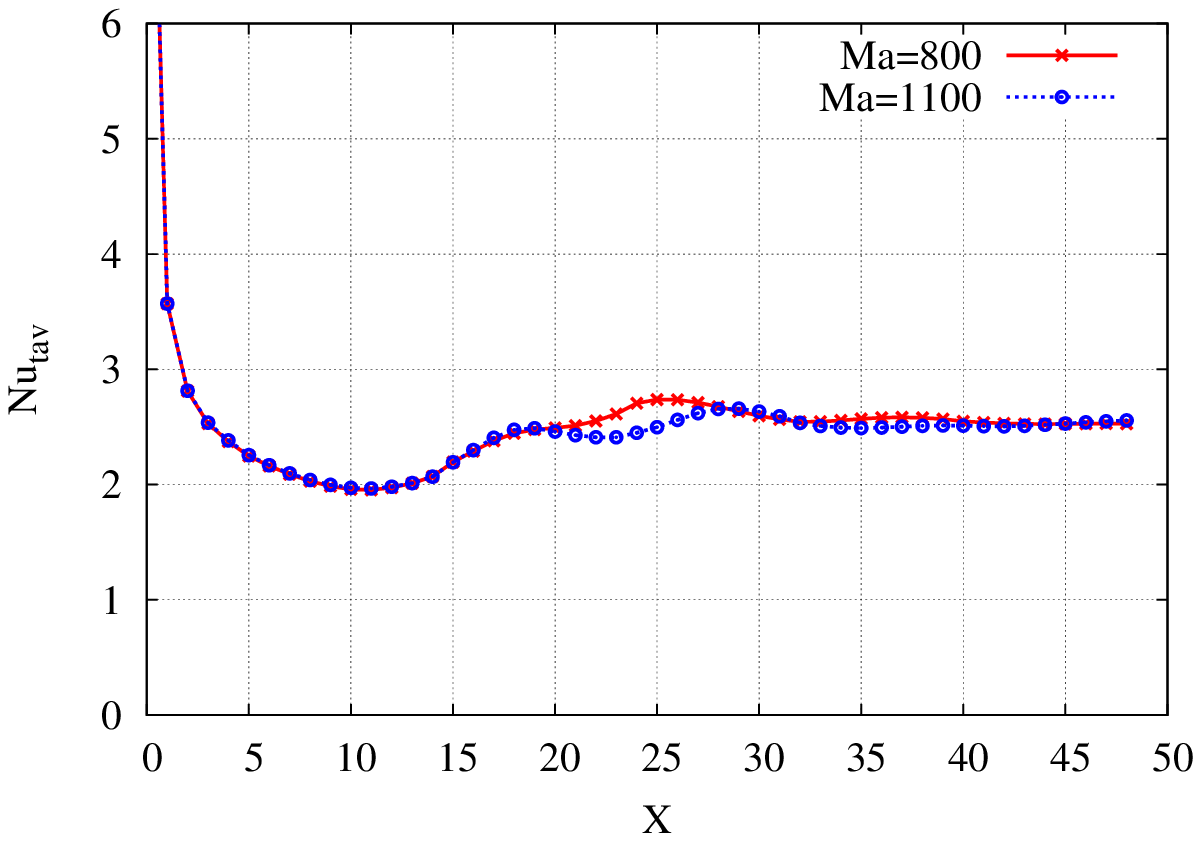}
\caption{\it \small The longitudinal variation of the transverse average Nusselt number for $Ma=800$ and $Ma=1100$. ($Ra=5000$, $Re=15$ and
$Bi=15$).}\label{Nut-ma}
\end{center}
\end{figure}

The variation of the surface tension with the temperature (Marangoni effect or thermocapillary effect) generates flows at the free surface from hot areas towards cold areas. Figure \ref{conv-ma} shows the effect of different Marangoni numbers on the temperature fields associated with the velocity vectors at the cross section ($X=40$) where the flow is fully developed by using different values of Marangoni number for $Ra=0$. Setting $Ra=0$ enables us to analyze the Marangoni effect only, without the interaction with buoyancy. We observe that the variation of the surface tension with temperature induces rolls similar to those created by natural convection. The results clearly indicate that, for $Ma=500$, the flow is dominated by forced convection. The first longitudinal vortex rolls start to appear at $Ma=1000$ where the thermocapillary effect on the basic flow becomes important. As the Marangoni number increases, more vortex rolls are induced and their number increases to reach their fully symmetrically developed state at $Ma=3000$. For this Ma value, the maximum number of vortex rolls is 8. The rolls' dimension is nearly the same as the channel height. The rotation direction of the rolls due to the Marangoni convection is reversed relative to the rotation direction of the rolls due to natural convection.\\
In the following section, we consider the effect of buoyancy thermal forces. Figure \ref{vy-ma} shows the transverse variation at $X=30$ of the transverse velocity component $V_y$ for various values of the Marangoni number $Ma$ and for $Re=15$ and $Ra=5000$. We observe that $V_y$ increases with increasing $Ma$, which confirms that the Marangoni effect induces a surface flow superimposed on the base flow. We noted also that for $Ma\geq1200$, there is a sign inversion of the velocity that is clearly visible for the rolls near the walls, which is due to the Marangoni effect overcoming the dominance of the buoyancy thermal forces. Figure \ref{effet-ma} presents a cross section of the flow at $X=30$ and shows the Marangoni effect on the longitudinal rolls. Under the influences of thermocapillary forces, the rolls near the side walls become larger when the Marangoni number increases, which induces the merging of the two center rolls in $Ma=1200$ and generates two other rolls at the sides of the channel. The two new rolls near the side walls then begin to grow with increasing $Ma$, causing the two center rolls to merge at $Ma=1800$. While the number of rolls decreases from 10 rolls at $Ma=1600$ to 8 rolls at $Ma=1800$, the rotational velocity of the rolls greatly increases and the rolls rotate in the direction imposed by the Marangoni effect. To study the influence of thermocapillary forces on the heat transfer, Figure \ref{nu-ma} shows the variation in the average Nusselt number at the heated bottom wall with respect to the Marangoni number. We note that the $Nu_{av}$ increases with increasing $Ma$ due to the increasing velocity of the rolls under the influence of thermocapillary forces (Figure \ref{vy-ma}). At $Ma=1100$, we observe a decrease in the Nusselt number due to the beginning of the merging of the two rolls in the channel center. This behavior is also shown by the evolution of the transverse average Nusselt number along the channel (Figure \ref{Nut-ma}). Indeed, the Nusselt number decreases initially and then increases gradually to a maximum that indicates the complete occupation of the cross section by the rolls, which occurs at $X=24$ for $Ma=800$ and at $X=28$ for $Ma=1100$. Thus, there is a downstream shift of the center rolls' position with respect to the outlet of the channel when $Ma$ increases between $800$ and $1100$. This decline continues as $Ma$ increases until the center rolls merge with their adjacent rolls at $Ma=1200$ (Figure \ref{effet-ma}). This reorganization of the rolls is also confirmed by the evolution of the velocity profile at the free surface for $Ma=1200$ (Figure \ref{vy-ma}). The increase in the rolls' velocity under the action of thermocapillary effects promotes the heat transfer and increases the Nusselt number until a further decrease occurs due to the disappearance of the two center rolls at $Ma=1800$.

\section{Conclusion}
This work presents a numerical study of three-dimensional laminar mixed convection within a liquid flowing in a horizontal channel heated uniformly from below. The upper surface is free and assumed to be flat. The side walls are thermally insulated. This numerical work allows us to examine the numerical results corresponding to a selected range of different characteristic non-dimensional parameters. We focused on the thermal instabilities that develop in the form of stationary longitudinal rolls. We investigated the influence of the variation of parameters governing the physics of the problem on the flow patterns and heat transfer characteristics.\\
The longitudinal rolls are first initiated near the side walls and grow gradually toward the center of the channel in the downstream direction. The rolls' dimension is nearly the same as the channel height. The Reynolds and the Rayleigh numbers have an important effect on the position of the rolls' development over the cross section. On the one hand, for a given Rayleigh number, the higher the Reynolds number, the greater the distance is at which the center rolls developed; on the other hand, for a given Reynolds number, higher Rayleigh numbers induce a lower distance at which the center rolls developed. Larger Reynolds and Rayleigh numbers favorably affected the heat transfer by forced convection in the case of a fixed Rayleigh number and the development of longitudinal rolls in the case of fixed Reynolds numbers. The presence of an upper wall (case of closed channel) increases the number of rolls compared with the free surface case. The imposed temperature $T_0$ at the top wall increases the thermal gradient in the film, which should improve the heat exchange. However, this favorable effect is counterbalanced by the viscous effects at this wall that decrease the fluid velocity and, consequently, decrease the heat exchange compared with the cases of $Bi=15$ and $Bi=20$. The results also show that the presence of the upper rigid wall stabilizes the flow, delaying the development of longitudinal rolls. Furthermore, the Marangoni effect has been investigated, and the results show that the thermocapillary effect on the basic flow becomes important at $Ma=1000$. As the Marangoni number increases, more vortex rolls are induced and increase their number to reach a maximum of 8 rolls. The rotation direction of the rolls due to the Marangoni convection is reversed relative to the rotational direction of the rolls due to natural convection. When we combine the Marangoni effect with the natural convection, we observe that the thermocapillary forces start to overcome the dominance of the buoyancy thermal forces at $Ma=1200$ for $Ra=5000$. Furthermore, increases in the Marangoni number favorably affect the heat transfer, but there are some drops, which corresponds to the disappearance of the center rolls.

\section* {Appendix A. Mesh size-dependence study}
Several grid sizes have been tested to determine the appropriate mesh size, giving the best compromise between accuracy and calculation cost to ensure that the results are grid-independent. The results are presented below for the case in which the control parameters are $Re=15$, $Ra=5000$, $Bi=15$ and $Ma=500$ for the Reynolds, Rayleigh, Biot and Marangoni numbers, respectively. Table \ref{tab:test} shows the average Nusselt numbers at the bottom surface of the channel for increasing meshes sizes. We note that the difference between the results of the last two meshes does not exceed 1 \textdiscount
\begin{table}[ht!]
\begin{center}
\begin{tabular}{|c|c|c|}
\hline
Meshes& Computation time (24 processor) & $Nu_{av}$ \\
\hline
102x140x40 & 2 h, 36min & 2.41663 \\
\hline
115x158x45 & 4 h, 41min & 2.43877 \\
\hline
128x176x50 & 12 h, 35min    & 2.47479 \\
\hline
153x211x60 & 51 h, 20min    & 2.47639 \\
\hline
\end{tabular}
\caption{\it \small A comparison of the average Nusselt number $Nu_{av}$ for various meshes }
\label{tab:test}
\end{center}
\end{table}

\begin{figure}[h!]
  \begin{center}
\includegraphics[width=0.60\textwidth]{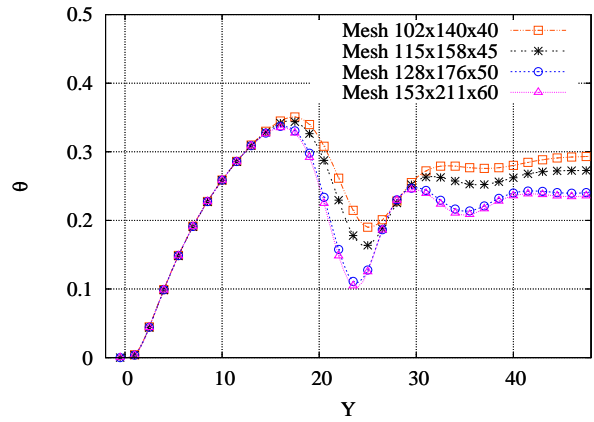} 
\caption{\it \small The longitudinal profile of reduced temperature along the axis $(Y, Z)=(5, 0.5)$ for different meshes.} \label{mesh}
  \end{center}
\end{figure}

Figure \ref{mesh} shows the longitudinal profile of the reduced temperature along the axis $(Y, Z)=(5, 0.5)$ for the four selected meshes. We observe that the results remain stable and are not subjected to any influence of mesh at a resolution of 128x176x50. Consequently, the resolution of the mesh that meets the right compromise between computational time and accuracy chosen for this study is 128x176x50.

\section*{Acknowledgments}

This work has been supported by a PBER grant (Programme de Bourses d'Excellence de Recherche) from CNRST-Morocco and by the Volubilis program of Morocco-French cooperation (MA/09/214-Maroc, MA/09/213-France). This work was performed using HPC resources from GENCI-CINES (Grant 2012-c201226738).

\section*{References}

\end{document}